\documentclass[prl, twocolumn, superscriptaddress, showpacs, floatfix, longbibliography]{revtex4}
\usepackage{graphicx,epsfig}% Include figure files
\usepackage{amsmath}% bold math
\usepackage{xcolor}% coloured text
\usepackage[breaklinks=true,colorlinks,citecolor=blue,linkcolor=blue,urlcolor=blue]{hyperref}
\def\be{\begin{equation}}
\def\ee{\end{equation}}

\def\ber{\begin{eqnarray}}
\def\eer{\end{eqnarray}}

\begin{document}
%

%\title{Plasmons in spin polarized graphene offer a new way to measure spin polarization}
\title{Plasmons in spin-polarized graphene: A way to measure spin polarization}
\author{Amit Agarwal}
\email{amitag@iitk.ac.in}
\affiliation{Department of Physics, Indian Institute of Technology Kanpur, Kanpur 208016, India}
\author{Giovanni Vignale}
\email{vignaleg@missouri.edu}
\affiliation{Department of Physics, University of Missouri, Columbia, Missouri 65211, USA}

\begin{abstract} 
We study the collective charge excitations (plasmons) in spin-polarized graphene, and derive explicit expressions for their dispersion in the  undamped regime. From this, we are able to calculate the critical  wave vector beyond which the plasmon enters the electron-hole continuum, its quality factor decreasing sharply. We find that the value of the critical wave vector is strongly spin polarization-dependent, in a way that has no analog in ordinary two-dimensional electron gases. The origin of this effect is in the coupling between the plasmon and the {\it inter-band} electron-hole pairs of the minority spin carriers. We show that the effect is robust with respect to the inclusion of disorder and we suggest that it  can be exploited to experimentally determine the spin polarization of graphene.
\end{abstract}
\maketitle
\section{Introduction}
Collective oscillations of the electronic charge (also known as plasmons), after being for many years a topic of fundamental research in solid state physics [\onlinecite{Pines, Giuliani_and_Vignale, Polini_rev, Stauber}], have recently emerged as a vector of information in a new generation of electronic devices, compressing the energy of electromagnetic field to the nanometer scale [\onlinecite{Maier,Yan, GTP}].  At the same time, a major effort is underway to couple the spin of electrons to electric fields with the objective to perform memory and logical operations.    In this context,  the interaction of plasmons with spins has received relatively little attention [\onlinecite{ag1,PhysRevLett.107.077004, ag2}].  It has generally  been assumed that spin polarization, occurring either spontaneously, or by the application of a magnetic field, or by spin injection under non equilibrium conditions, would have only a minor effect on the propagation of the charge mode. 
In a recent paper, we have pointed out the possibility that a new collective mode -- a spin mode -- might appear in a strongly spin-polarized Galilean-invariant  two-dimensional electron gas (2DEG) within the pseudo-gap that separates the high-frequency single-particle excitations  of the majority-spin electrons from the low-frequency single-particle excitations of the minority-spin component [\onlinecite{ag2}].  While this is a qualitatively new feature of spin-polarized electron liquids, it does not affect the main characteristics of the charge plasmon.
   
The situation is profoundly different and far more interesting in the two-dimensional spin-polarized  electron gas in graphene [\onlinecite{Shung, Guinea_NJP2007, SDS1, Polini1, Pyatkovskiy}], which is the subject of this paper.   At the root of the difference is the existence of low-energy inter-band excitations in which an electron from the lower cone of the massless Dirac fermion band structure is promoted to the upper cone without spin flip.  These inter-band excitations are  absent in the ordinary Galilean-invariant 2DEG.  As the spin polarization of the electron gas increases, the energy of inter-band excitations of the minority-spin component decreases and eventually it becomes so small that it matches, at a critical wave vector $q_c$, the energy of the charged plasmon at the same wave vector.   When this happens, the plasmon ceases to be a well-defined excitation, since there is nothing to prevent it from decaying into the inter-band electron-hole  (e-h) pairs of the minority-spin component (Landau damping).  
Experimentally, the onset of this decay process should show up as a sharp drop in the value of the quality factor, that is to say, the number of plasma oscillations that occur  during an average plasmon lifetime.  This sharp onset, being strongly spin-dependent, offers a new and unexpected way to determine the spin polarization of electrons in graphene from measurements of the plasmon lifetime.

This paper is organized as follows. In Sec.~\ref{sect:RPAtheory} we review the RPA theory of plasmons in clean spin polarized graphene.  We calculate the undamped plasmon dispersion as a function of spin polarization and show that, at variance with the Galilean-invariant 2DEG, there can be no spin plasmon for any spin polarization, because the Fermi velocities of up- and down-spin electrons are identical.  We go on to compute the onset of Landau damping of the charge plasmon and the spin polarization dependence of the quality factor of this mode. 
In Sec.~\ref{Sec3}, we show that our results are robust against the inclusion of disorder.  Specifically, we show that short-range impurities in a high-mobility sample introduce a lower wave vector $q_c^*$  below which the charge plasmon becomes unstable to diffusion.  Above this wave vector the plasmon dispersion calculated in Sec.~\ref{sect:RPAtheory} is essentially unchanged, and so is the sharp onset of Landau damping at $q_c$. Section~\ref{Sec4} summarizes our results and conclusions.        

\section{RPA theory of plasmons in Graphene}
\label{sect:RPAtheory}
We consider doped monolayer graphene at  zero-temperature, within the regime of applicability of the low energy linear-band dispersion relation [\onlinecite{Neto}], with a finite spin polarization. Spin polarization in graphene can be induced through various means, such as  non-equilibrium spin injection [\onlinecite{Han,Kamalakar}], or optical excitation [\onlinecite{Bottegoni}], or proximity effect [\onlinecite{Proxymity}]. 
However, for the sake of modeling spin-polarized graphene in the simplest possible way,  we assume that the polarization is induced by an in-plane Zeeman field $B$,  
which has negligible orbital effects. 
For a given valley, the energy dispersion of the spin-resolved bands is given by 
\be 
\varepsilon_{\xi \sigma}(k) = \xi~ \hbar v_{\rm F} k  + \sigma g_{\rm b} \mu_{\rm B} B~,  
\ee 
where $\xi = +~(-)$ indicates the conduction (valence) band, $\sigma = +~(-)$ indicates $\uparrow$-spin ($\downarrow$-spin)  electrons, $v_{\rm F}$ is the Fermi velocity in graphene, which is typically $10^6 $m/s --- independent of the applied Zeeman field,
and $g_{\rm b}$ is  the Land\'{e} $g$-factor,  while $\mu_{\rm b}$ is the Bohr magneton. The total density of states in graphene is given by $ g_s g_v |\varepsilon|/(2 \pi v_{\rm F}^2)$, with $g_s =2 $ and $g_v =2 $ being the spin and valley degeneracies, respectively. The Fermi momentum in unpolarized graphene is $k_{\rm F} \equiv \sqrt{ 4 \pi n /g_sg_v} $, and the associated Fermi energy is $\varepsilon_{\rm F} = \hbar v_{\rm F} k_{\rm F}$.
The spin polarization is  defined as $P\equiv (n_\uparrow - n_\downarrow)/(n_\uparrow + n_\downarrow) \approx -sgn(B) (g_{\rm b} \mu_{\rm b} B/\varepsilon_{\rm F})^2$ for small magnetic fields, where $n_{\sigma}$ is the spin polarized carrier density in graphene.  The spin-resolved Fermi wave vectors $k_{\rm F \sigma}$ can be expressed in terms of $P$ as $k_{\rm F \sigma} = k_{\rm F} \sqrt{1+ \sigma P} $. In the rest of the paper we will consider $P> 1$ without loss of generality. 
The Wigner-Seitz parameter $r_s$, which indicates the relative strength of Coulomb interactions,  is a constant  and it is given by $r_s = e^2/ \kappa \hbar v_{\rm F}$, where $\kappa$ is the static dielectric constant of the substrate hosting graphene monolayer.

Following Refs.~[\onlinecite{Giuliani_and_Vignale, ag2}] the collective density excitations within RPA in a spin-polarized medium are given by the zeros of the complex longitudinal dielectric function 
\ber \label{eq:eps}
\epsilon(q, \omega) &\equiv&  1- V_q [\chi_{0 \uparrow}(q,\omega) + \chi_{0 \downarrow}(q,\omega)]~ \nonumber \\ 
&=& 0~,
\eer
where $\chi_{0 \sigma}$ are the spin-resolved response functions of the non-interacting spin-polarized medium, and $V_q$ denotes the Fourier transform of the screened or unscreened Coulomb potential.  
The oscillation frequency of various collective modes $\omega_{\rm osc}$ (typically charge and spin plasmons) and the corresponding damping rate (or inverse lifetime) $\gamma_{\rm osc}$,  is obtained by solving for the complex roots of Eq.~\eqref{eq:eps}:  $\omega = \omega_{\rm osc}(q) - i \gamma_{\rm osc}(q)$, for a given $q$.  
Note that for the stability of the collective mode it is essential that $\gamma_{\rm osc}>0$. The collective mode typically decays by creating single electron-hole pairs (Landau damping), which occurs if the frequency of the collective mode lies in the e-h  continuum, which in graphene can arise either from intra-band or inter-band e-h excitations. 
%
%%%%%%%%%%%% 
\begin{figure}[t]
\begin{center}
\includegraphics[width=1.0 \linewidth]{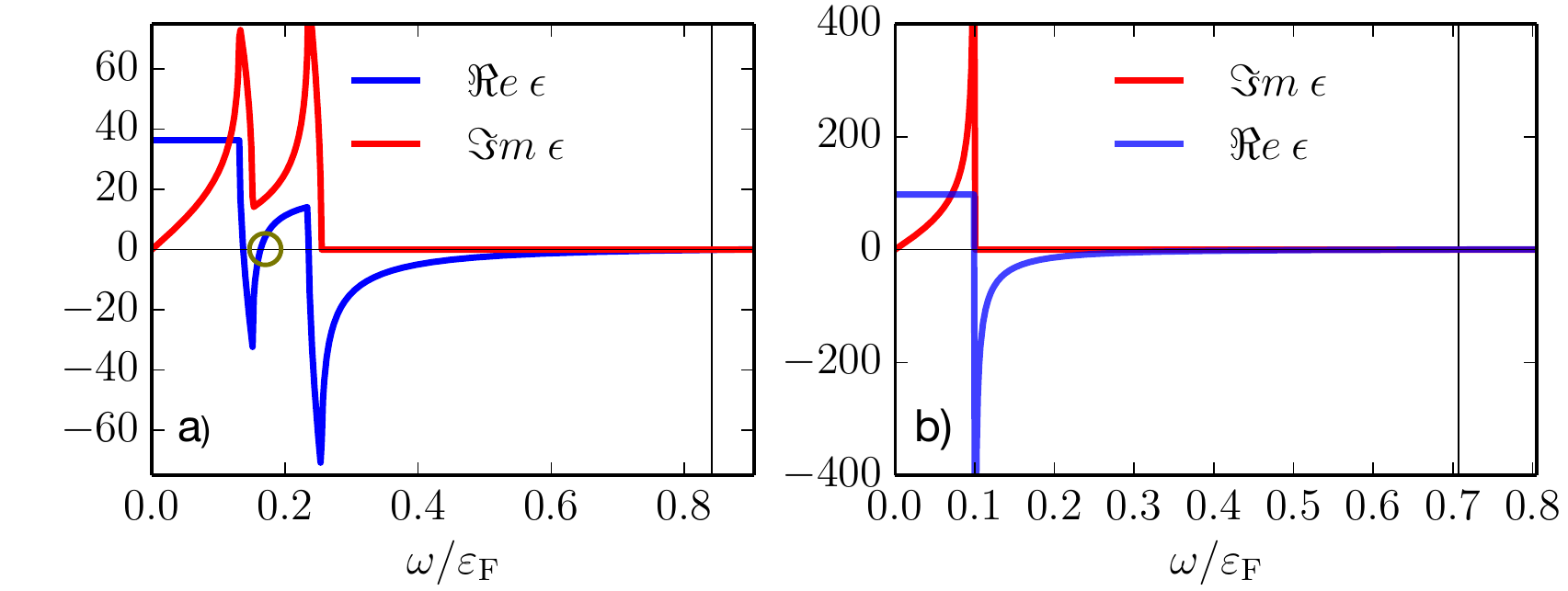}
\caption{ (a) Real and imaginary parts of the RPA dielectric function $\epsilon(q,\omega)$ vs $\omega$ for a 2DEG (parabolic dispersion) system. The zero of the dielectric function marked by the small circle corresponds to the spin plasmon mode in 2DEG \cite{ag2}.  (b) Real and imaginary parts $\epsilon(q,\omega)$ vs $\omega$ for graphene. Note that there is no spin plasmon mode in graphene.  Other parameters are chosen to be $P = 0.5$, $q/k_{\rm F}= 0.1$, and $r_s =2.5$ for both panels. The thin vertical lines in both panels mark the location of the known plasmon mode in the long wavelength limit, i.e.,  in graphene $\omega_{\rm pl}/\varepsilon_{\rm F} = \sqrt{2r_sq/k_{\rm F}}$ and in 2DEG $\omega_{\rm pl}/\varepsilon_{\rm F} = \sqrt{2^{3/2} r_s q/k_{\rm F}}$. 
\label{fig:fig1}}
\end{center}
\end{figure}
%%%%%%%%%%%%%%%%%%%%%%%%%%%%%%%%%%%%%%%%%%%%%%%%%%%
%
If the damping rate $\gamma_{\rm osc}$ is small, it can be estimated by doing a Laurent-Taylor  expansion of the dielectric function around $\omega_{\rm osc}$ [\onlinecite{ag2}]. For a frequency independent $V_q$, it is given by 
\be \label{eq:gamma}
\gamma_{\rm osc}(q) =  \left. \frac{\Im m [\epsilon(q,\omega)]}{\partial \Re e[\epsilon (q,\omega)]/\partial \omega} \right|_{\omega = \omega_{\rm osc}}  ~.    
\ee
In an actual experimental scenario, the physical observability of a damped collective mode depends on the sharpness of its resonance peak, which is captured by the quality factor of the collective mode, defined as  $Q = \omega_{\rm osc}(q)/\gamma_{\rm osc}(q)$. 

The spin-polarized Lindhard function for graphene [\onlinecite{Guinea_NJP2007}] is best expressed in terms of the dimensionless variables, 
\be \label{eq: nupm}
\nu_{\pm \sigma} = 2 \frac{k_{\rm F \sigma}}{q} \pm \frac{\omega}{v_{\rm F} q}~, 
\ee
and at $T=0$, for complex frequencies it is explicitly given by [\onlinecite{Pyatkovskiy, Page}]
\be
\frac {\chi_\sigma (q,\omega)}{N_0} = -\frac{k_{{\rm F}\sigma}}{k_{\rm F}} + F(q, \omega) \Big[ G_+(\nu_{+ \sigma}) + G_-(\nu_{-\sigma})\Big]~,
\ee
where
\be 
F(q,\omega) = \frac{v_{\rm F} q^2}{8 k_{\rm F} \sqrt{v_{\rm F}^2 q^2- \omega^2}}~,
\ee
and 
\be \label{eq:G}
G_{\pm}(z) = z \sqrt{1-z^2} \pm i \cosh^{-1}(z)~.
\ee

In the  long wavelength limit,  $q \to 0$ with $\omega > v_{\rm F} q$ fixed, which is important for optical spectroscopy and the plasma oscillations, the spin-polarized Lindhard function can be approximated as 
\be \label{eq:chi_approx}
\frac{\chi_{0 \sigma}}{N_0} \approx (1+ \sigma P)^{1/2} \left( -1 + \frac{\omega}{\sqrt{\omega^2 - v
_{\rm F}^2 q^2}}~\right).
\ee
As a check of Eq.~\eqref{eq:chi_approx} we note that a similar approximate expression for the Lindhard function (without spin polarization) has also been derived in the context of topological insulators, [\onlinecite{RaghuPRL2009}] which have the same quasiparticle dispersion as graphene.

Having calculated the Lindhard function we now focus on the zeros of the dielectric function, i.e., the collective modes.
In Fig.~1, we plot the dielectric function of graphene [panel (b)] and compare it with the same for 2DEG [panel (a)]. Evidently, there are two stable collective modes in 2DEG, corresponding to charge and spin density excitations [\onlinecite{ag2}], while in graphene there is only one stable collective mode and it corresponds to charge plasmons. Physically this is a direct consequence of the fact that in a 2DEG with parabolic dispersion relation, the Fermi velocity of the majority spin carriers can be significantly larger than the Fermi velocity of the minority spin carriers.   This opens up a region of reduced spectral density (pseudogap), which is clearly visible in Fig.~\ref{fig:fig1}a) between the peaks associated with e-h excitations of the majority and minority spin carriers: it is in this window that the spin plasmon lives.  At variance with this,  the Fermi velocities of graphene  are the same for both spin species and, as a consequence, there is no spectral window in which the spin plasmon can find a home (see Fig.~\ref{fig:fig1}).

We now proceed to calculate the dispersion of the charge plasmon in spin-polarized graphene [\onlinecite{Shung, Guinea_NJP2007, SDS1, Polini1}]. 
Substituting Eq.~\eqref{eq:chi_approx} in Eq.~\eqref{eq:eps} leads to the following approximate plasmon dispersion, 
\be \label{eq:pl}
%\omega_{\rm pl}^2 =  v_{\rm F}^2 q^2   \frac{\left(1+  V_q N_0 \delta \right)^2}{1+ 2V_q N_0 \delta }~, 
\omega_{\rm pl}(q) =  v_{\rm F} q    \frac{1+  V_q N_0 \delta(P)}{\sqrt{1+ 2V_q N_0 \delta(P)} }~, 
\ee
where we have defined $\delta (P) \equiv \sqrt{1+P} +\sqrt{1-P}$. Note that $\delta(P)$ is a monotonically decreasing function of  $P$, varying from $\delta(0) =2$ to $\delta(1) = \sqrt{2}$. 
 For the usual case of long range unscreened Coulomb repulsion in 2D, $V_q = 2 \pi e^2/\kappa q$, where $\kappa$ is the dielectric constant of the substrate,  we have $N_0 V_q = 2 r_s k_{\rm F}/q$ (with the factor of two originating from the valley degeneracy) and Eq.~\eqref{eq:pl} can be rewritten as, 
\be \label{eq:pl2}
%\omega_{\rm pl}^2 =  v_{\rm F}^2 q ~ \frac{\left(2 \delta r_s k_{\rm F} + q \right)^2}{4 \delta r_s k_{\rm F}  + q }~.
\omega_{\rm pl}(q) =  v_{\rm F} \sqrt{q} ~ \frac{2 \delta(P) r_s k_{\rm F} + q }{\sqrt{4 \delta(P) r_s k_{\rm F}  + q} }~.
\ee
The plasmon mode in  spin polarized graphene enters the e-h continuum of the inter-band transitions of the minority spin species when it crosses the region specified by $ \omega + v_{\rm F} q \geq 2 k_{\rm F} v_{\rm F} (1-P)^{1/2}$,  and the critical wave vector is obtained to be 
\be \label{eq:qcP}
q_c (P) = \frac{A(P) + \sqrt{A(P)^2 + 16 k_{\rm F}^2 r_s \delta(P) (1-P)^{3/2} } }{2 \sqrt{1-P}}~,
\ee 
where $A(P) \equiv k_{\rm F} [(1-P)  -4 r_s \delta(P) \sqrt{1-P} - r_s^2 \delta(P)^2] $. In the limiting case of $P \to 0$, i.e., unpolarized graphene, Eq.~\eqref{eq:qcP} reduces to 
\be \label{eq:qc0}
q_c (0) = \frac{ k{\rm _F}}{2} \left[ 1-8r_s -4 r_s^2 + (1+2 r_s) \sqrt{ 1+12 r_s + 4 r_s^2} \right]~.
\ee
As a consistency check we note that Eq.~\eqref{eq:pl2} can be expanded around $q \to 0$ to obtain
\be
%\omega_{\rm pl}^2 \approx    r_s k_{\rm F} \delta(P) ~ v_{\rm F}^2 q + \frac{3}{4} v_{\rm F}^2 q^2~. 
\frac{\omega_{\rm pl}(q)}{k_{\rm F} v_{\rm F}} \approx    \sqrt{r_s \delta(P)}\left(\frac{q}{k_{\rm F}}\right)^{1/2} + \frac{3}{8\sqrt{r_s \delta(P)}}  \left(\frac{q}{k_{\rm F}}\right)^{3/2}~. \label{eq:pl3}
\ee
Here the first term in Eq.~\eqref{eq:pl3} gives the well known $\sqrt{q}$ behavior of the plasmon dispersion in 2D, and the second term, which also depends on the spin polarization,  gives the additional correction which was reported earlier in the context of plasmons in unpolarized ($P=0,~ \delta(P) =2$) intercalated graphite [\onlinecite{Shung}]. 

\begin{figure}[t!]
\begin{center}
\includegraphics[width=1.0 \linewidth]{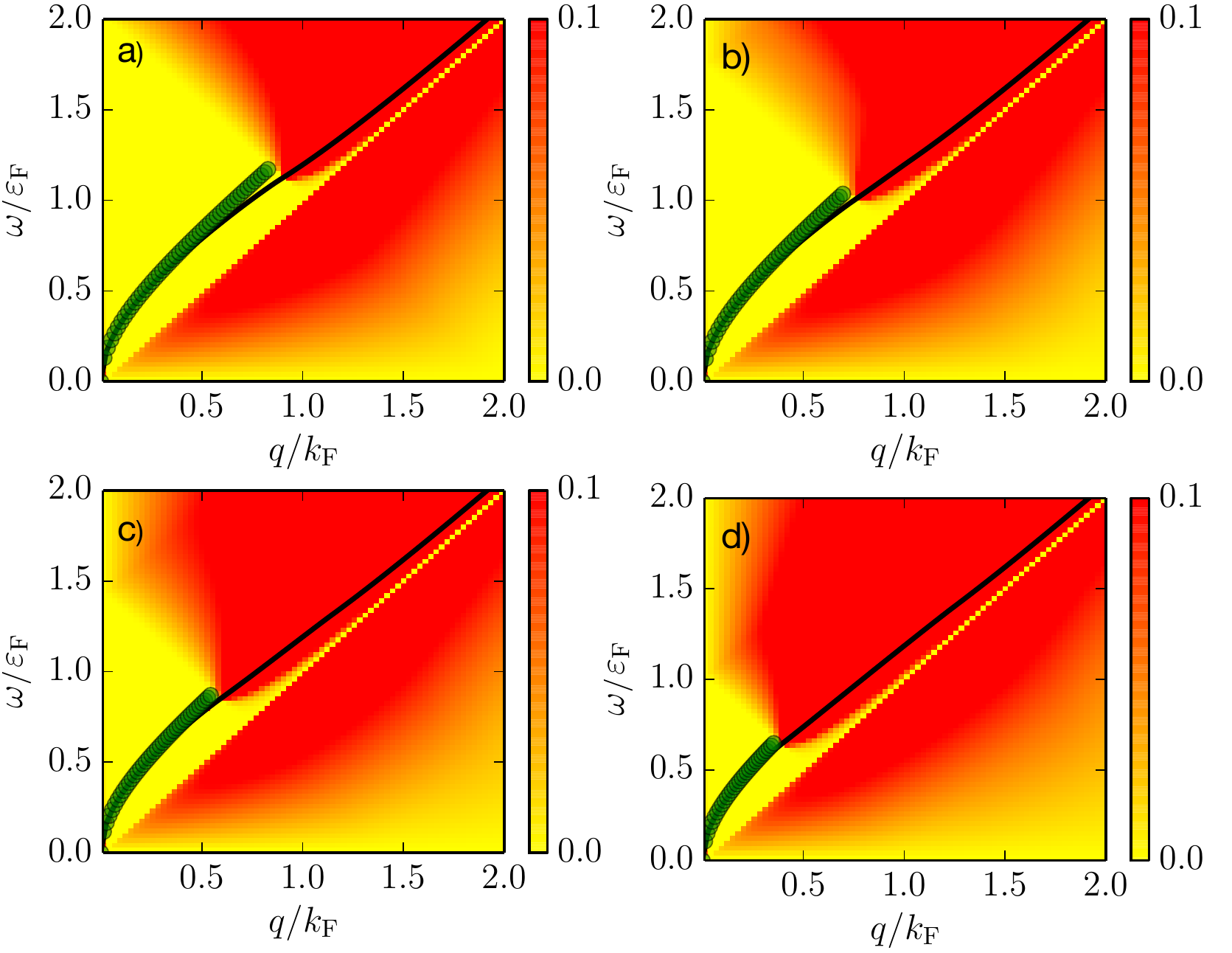}
\caption{  The color plot of the loss function, $-\Im m [1/\epsilon(q,\omega)]$, 
along with the numerically evaluated plasmon dispersion (solid black line), and the approximate analytical dispersion (green circles) as per Eq.~\eqref{eq:pl2}.  Panels (a), (b), (c), and (d) correspond to $P=0,~0.25, ~0.5,~ \rm{and} ~0.75$ respectively.   Here $r_s = 0.5$ in all the panels. 
\label{fig:fig2}}
\end{center}
\end{figure}

\begin{figure}[t]
\begin{center}
\includegraphics[width=.99 \linewidth]{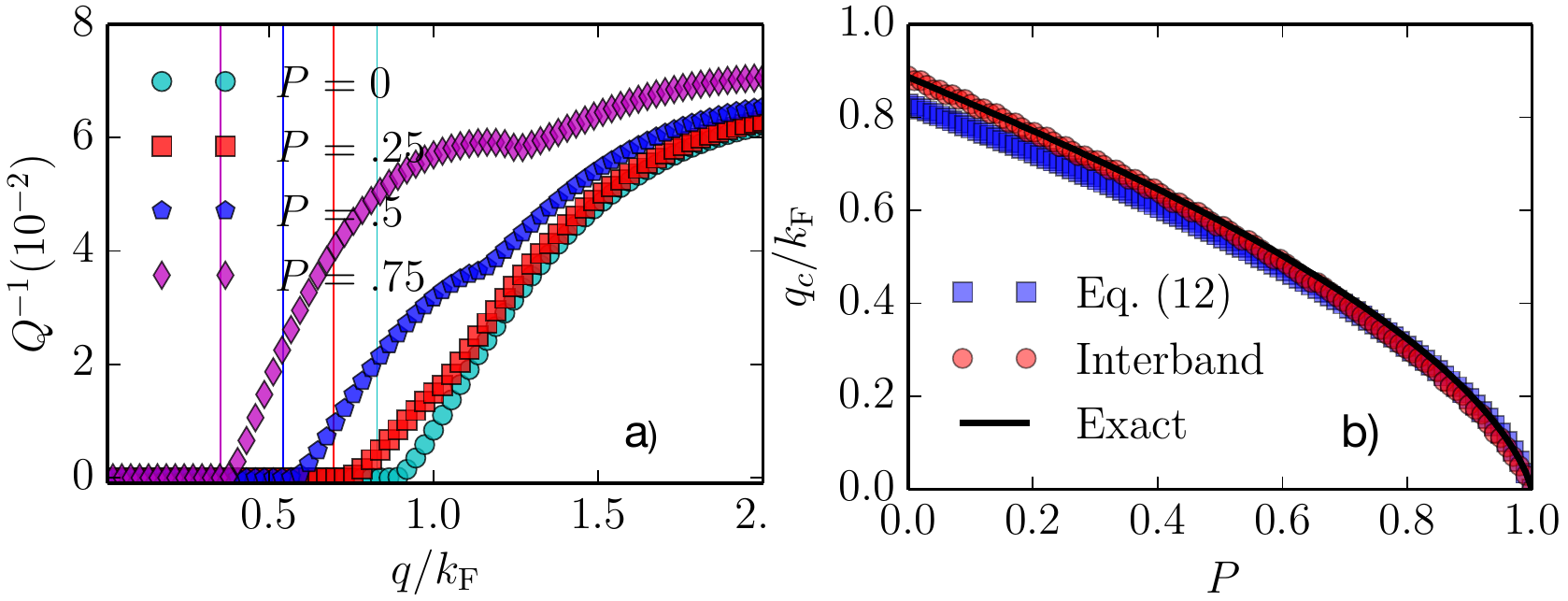}
\caption{(a) $Q^{-1}$ vs $q$ for various values of $P$ in the clean limit. A sharp rise occurs when the charge plasmon enters the inter-band e-h continuum of the minority spin species and is approximately given by Eq.~\eqref{eq:qcP} and marked by thin vertical lines. A second cusp occurs when the plasmon enters the  inter-band e-h continuum of the majority spin species, whose approximate location is specified by Eq.~\eqref{eq:qcP} with $P \to -P$, and becomes visible for large values of $P$. (b) shows the critical $q_c$ beyond which the plasmon mode enters the inter-band  e-h continuum.  The $P$ dependence of the critical $q_c$ arises almost entirely from the shift in the boundary of the inter-band continuum:  this is shown by the curve labeled ``interband", in which the polarization dependence of the plasmon dispersion has been neglected.  Further the maximum deviation of Eq.~\eqref{eq:qcP} from the exactly calculated critical wave vector occurs for $P \to 0$, and is less that $6.7 \%$.  We have taken $r_s = 0.5$ in both the panels. 
\label{fig:fig3}}
\end{center}
\end{figure}

In Fig.~\ref{fig:fig2} we plot the loss function and the plasmon dispersion, both numerically calculated (solid black curve) and evaluated from the analytical expression~\eqref{eq:pl2}  (green circles) for $P=0,~0.25, ~0.5,~\rm{and} ~0.75$. Observe how increasing the degree of spin polarization leads to a decreasing value of $q_c$, the critical wave vector for which the plasmon mode enters the inter-band continuum of the minority spin species, i.e., $\omega_{\rm pl}(q_c) = 2 \hbar v_{\rm F} k_{{\rm F}\downarrow} - v_{\rm F} q_c$ for $P>0$.   This is also reflected in the inverse of the quality factor as a function of $q$, which should be zero for $q < q_{\rm c}$ since the mode is completely undamped in the present approximation, and will become finite as soon as the plasmon mode enters the inter-band e-h continuum of the minority spin carriers. This is clearly shown in Fig.~\ref{fig:fig3} which also highlights the fact that the critical $q_{\rm c}$ decreases with increasing spin polarization. This jump in the quality factor of the plasmon mode at $q= q_{\rm c}$ can be used to estimate $q_{\rm c}$ experimentally,  and comparing this  to either Eq.~\eqref{eq:qcP} or the exact numerical value of $q_{\rm c}$, should give a good estimate of the degree of spin polarization $P$. 

\section{Effect of diffusion on the plasmon dispersion} 
\label{Sec3}
In this section we study the effect of weak disorder on the charge plasmon mode [\onlinecite{giuliani_prb_1984}] in spin-polarized graphene. For sufficiently dilute impurities, the collective modes of the disordered system are still given by Eq.~(\ref{eq:eps}), provided that the bare Lindhard function is replaced  by the disorder-averaged response function. A suitable expression for the latter was given by Mermin [\onlinecite{mermin_prb_1970}], based on a relaxation time approximation. It is given by 
\be \label{eq: mermin}
\Pi_{0\sigma} (q, \omega) = \frac{ (\omega + i \tau^{-1}) ~\chi_{0\sigma}( q, \omega+  i/\tau)}{\omega + i \tau^{-1} \chi_{0\sigma}(q, \omega+ i/\tau)/\chi_{0\sigma} (q,0)}~,
\ee
where $\chi_{0\sigma}(q,\omega)$ is the Lindhard response function of the clean system, and $\tau$ is the elastic lifetime of the momentum eigenstates in the presence of static disorder. 
Note  that Eq.~\eqref{eq: mermin} is  actually equivalent to the sum of impurity ladder diagrams (diffusons) in the diffusive regime [\onlinecite{giuliani_prb_1984}], and it also has the correct high-frequency behaviour (the collisionless regime).

\begin{figure}[t!]
\begin{center}
\includegraphics[width=.99 \linewidth]{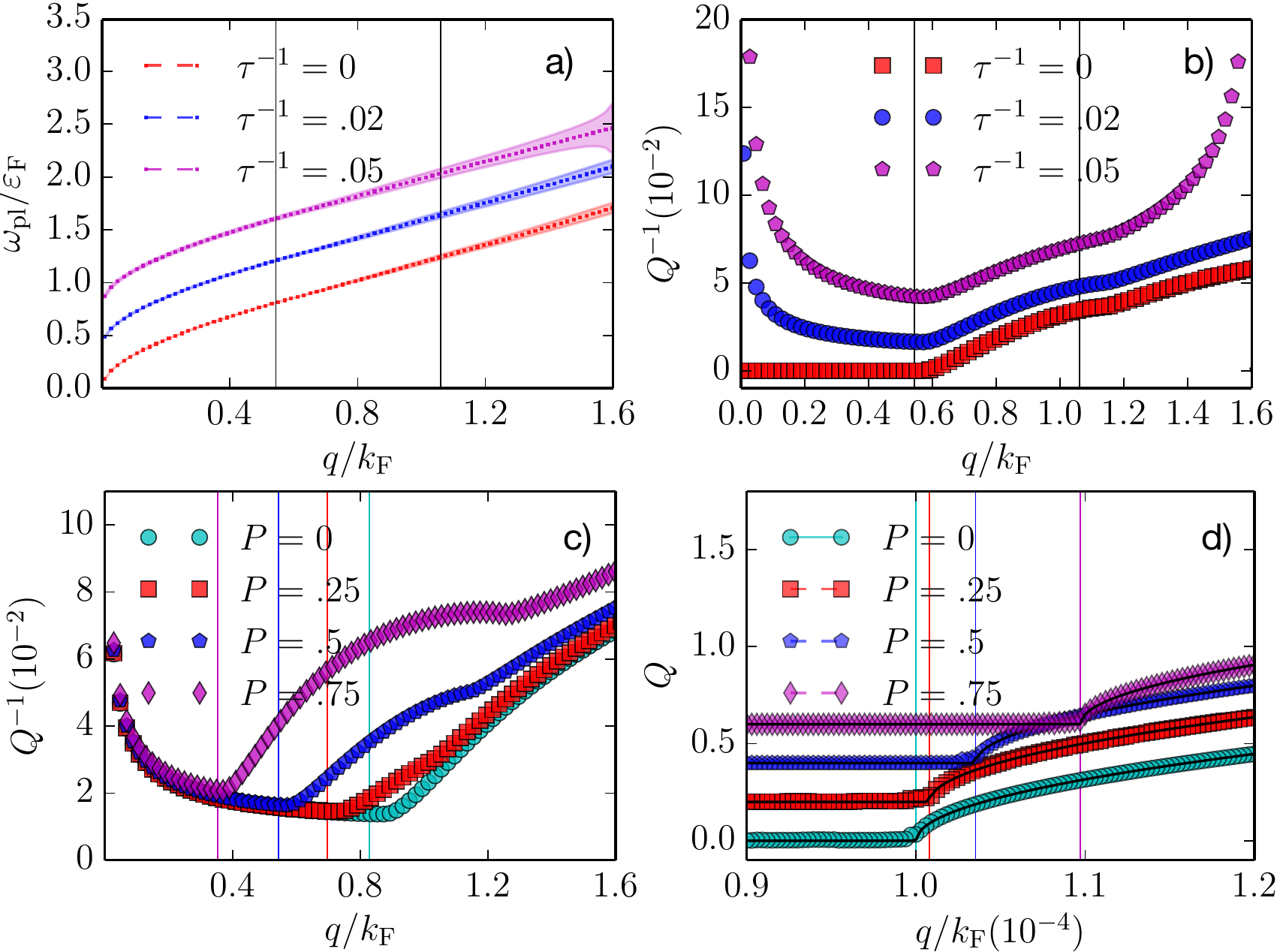}
\caption{(a)  and (b) display $\omega_{\rm pl}$, and $Q^{-1}$, respectively, as functions of $q$ for different disorder strengths, for $P=0.5$. The shaded region around $\omega_{\rm pl}$ indicates the decay rate $\gamma_{\rm pl}$, and the plasmon curves for different disorder strength are shifted vertically (by $0.4$) for better visibility. (c) and (d) plot  $Q^{-1}$ for different polarization strengths, with $1/\tau = 0.02$, for large- and small $q$, respectively.  In (c) the cusps in the inverse quality factor arise for the critical wave vectors where the plasmon mode enters the {\it inter-band} e-h continuum of the minority (first cusp -- marked by vertical lines), and then the majority spin carriers (second cusp).  (d) highlights the existence of the $P$-dependent critical wave vector of Eq. \eqref{eq:qstar}, below which the plasmon modes are unstable to diffusion.  Notice that we plot $Q$ rather than $Q^{-1}$ and the different curves are shifted vertically (by $0.2$) for better visibility.  The symbols indicate the exact numerical result, which are in excellent agreement with the approximate analytical result of Eq.~\eqref{eq:Q} indicated by the solid black line. In all panels, we have taken $r_s = 0.5$.
\label{fig:fig4}}
\end{center}
\end{figure}

An approximate disorder-averaged Lindhard function can now be obtained by substituting Eq.~\eqref{eq:chi_approx} in Eq.~\eqref{eq: mermin}, yielding
\be \label{eq:Pidis}
\frac{\Pi_{0\sigma}}{N_0}  \approx \sqrt{1+ \sigma P} \left(- 1 + \frac{\omega}{[(\omega + i /\tau)^{2} - v_{\rm F}^2 q^2]^{1/2} - i /\tau} \right) ~.
\ee
Note that, except for the spin dependent pre factors, Eq.~\eqref{eq:Pidis} has a functional form similar to that of the disordered response function of a 2DEG with parabolic dispersion relation, evaluated by means of a perturbation theory in Ref.~[\onlinecite{giuliani_prb_1984}].  Now even the low energy plasmon dispersion will have an imaginary component and can be obtained by substituting Eq.~\eqref{eq:Pidis} in Eq.~\eqref{eq:eps}. It is explicitly given by 

\be \label{eq:pldis1}
\omega_{\rm pl} = \frac{1 + V_q N_0 \delta}{1 +  2 V_q N_0 \delta} \left[\sqrt{v_{\rm F}^2 q^2 [1 +  2 V_q N_0 \delta] - 1/\tau^2} - i/\tau \right]~,
\ee
where $\delta(P)$ is defined below Eq.~\eqref{eq:pl}. For unscreened Coulomb repulsion $V_q = 2 \pi e^2/q$, Eq.~\eqref{eq:pldis1} reduces to 
\be \label{eq:pldis2}
\omega_{\rm pl} = \frac{q + 2 r_s k_{\rm F} \delta}{q + 4 r_s k_{\rm F} \delta}  \left[\sqrt{v_{\rm F}^2 q [q + 4 r_s k_{\rm F} \delta] - 1/\tau^2} - i/\tau \right]~.
\ee
Note that Eq.~\eqref{eq:pldis2} reduces to Eq.~\eqref{eq:pl2} in the limiting case of vanishing disorder $1/\tau \to 0$. 

Equation~\eqref{eq:pldis2} immediately implies that in spin-polarized graphene with dilute disorder, collective density excitations can exist only for wave-vectors greater than a critical wave-vector ($q^*_c$) which is analytically given by 
\be \label{eq:qstar}
q^{*}_c = -2 r_s k_{\rm F} \delta(P)  +  \sqrt{(2 r_s k_{\rm F} \delta(P))^2 +  1/(v \tau)^2}~.
\ee
Equation~\eqref{eq:qstar} is consistent with the physical intuition that coherent collective excitations cannot exist for length scales longer than the mean free path, i.e., long-wavelength plasmons ``diffuse" in the disordered medium.  
The approximate value of the quality factor (for $q>q_c^*$) can be obtained from Eq.~\eqref{eq:pldis2}, and is given by 
\be \label{eq:Q}
Q = \tau \sqrt{v_{\rm F}^2 q (q + 4 r_s k_{\rm F} \delta(P)) - 1/\tau^2}~.
\ee
In panels b) and c) of Fig.~\ref{fig:fig4} we plot the inverse quality factor $Q^{-1}$ vs $q$ for various strengths of disorder (b) and degrees of polarization (c).   The overall behavior of $Q^{-1}$ is a U-shaped curve -- the quality factor decreasing sharply both at large wave vectors (where Landau damping from inter-band excitations begins) and at low wave vector, where diffusion damping takes over.   In the region of very small wave vectors, a plot of the quality factor itself (panel d) more conveniently illustrates the strong damping of the plasmon below the lower critical wave vector $q_c^*$. The  dependence of the lower 
critical wave vector $q_c^*$  on spin polarization could also be used, in principle, for an experimental estimation of $P$, but this approach would suffer from much larger uncertainties than the one based on the determination of the upper critical wave vector $q_c$.

\section{Conclusion}
\label{Sec4}
In summary,  we have studied the collective density excitations of spin-polarized graphene, and derived explicit expressions for their dispersion and the critical wave vector for the onset on Landau damping, considering the clean case as well as the weakly disordered case. Our analytical results are in excellent agreement with the exact numerical results. We have found that, unlike the case of 2DEG,  both the plasmon dispersion, and the upper critical wave-vector beyond which the plasmon mode decays into the e-h continuum, are strongly spin polarization dependent. Further we have shown that this effect is robust with respect to the inclusion of disorder. The dependence of the upper critical wave-vector on spin polarization will manifest as a cusp in the measured quality factor, which we propose can be used to experimentally determine the spin polarization of graphene.
  
\section*{Acknowledgements}  
We  acknowledge funding support from the DST INSPIRE Faculty Award (AA) and 
 from the NSF Grant No. DMR-1406568 (GV).

%\bibliography{SPGraphene}

%
\end{document}